\documentclass[10pt,final,twocolumn,journal]{IEEEtran}
\usepackage{url}
\usepackage{graphicx}
\usepackage[noadjust]{cite}
\usepackage{epstopdf}
\usepackage{cite}
\usepackage{caption}
\usepackage{subcaption}
\usepackage{authblk}
\usepackage{xcolor}
\usepackage{amsmath}
\usepackage{array}
\newcolumntype{P}[1]{>{\centering\arraybackslash}p{#1}}
\usepackage{enumitem}
\usepackage{booktabs}
\usepackage{adjustbox}
\usepackage[font=small,labelfont=bf]{caption}
\begin{document}

\title{Impact of body thickness and scattering on III-V triple heterojunction Fin-TFET
modeled with atomistic mode space approximation}

\author{Chin-Yi Chen, Hesameddin Ilatikhameneh, Jun Z. Huang, Gerhard Klimeck, Michael Povolotskyi\vspace{-6.5ex}
\thanks{This was supported by National Science Foundation E2CDA Type I collaborative research on "A Fast 70mV Transistor Technology for Ultra-Low-Energy Computing" with the award number of 1639958 and semiconductor research corporation with its task ID of 2694.003. The use of nanoHUB.org computational resources operated by the Network for Computational Nanotechnology funded by the US National Science Foundation under Grant Nos. EEC-1227110, EEC-0228390, EEC-0634750, OCI-0438246, and OCI-0721680 is gratefully acknowledged. NEMO5 developments were critically supported by an NSF Peta-Apps award OCI-0749140 and by Intel Corp. This work also used the Extreme Science and Engineering Discovery Environment (XSEDE) at SDSC Dell Cluster with Intel Haswell Processors (Comet) through 50,000.0 SUs under charge number TG-ECS190009. And, we greatly appreciate Prof. Mark Rodwell, Hsin-Ying Tseng, and Tarek A. Ameen's constructive discussions.}
\thanks{The authors are with the Department of Electrical and Computer Engineering, Purdue University, West Lafayette, IN, 47907 USA e-mail: r99941001@gmail.com}
}
\maketitle

\setlength{\textfloatsep}{12pt}
\setlength{\belowdisplayskip}{1.6pt} 
\setlength{\belowdisplayshortskip}{1.6pt}
\setlength{\abovedisplayskip}{1.6pt} 
\setlength{\abovedisplayshortskip}{1.6pt}
\setlength{\belowcaptionskip}{-12pt}
\vspace{-1.0\baselineskip}
\begin{abstract}
The triple heterojunction TFET has been originally proposed to resolve TFET's low ON-current challenge. The carrier transport in such devices is complicated due to the presence of quantum wells and strong scattering. Hence, the full band atomistic NEGF approach, including scattering, is required to model the carrier transport accurately. However, such simulations for devices with realistic dimensions are computationally unfeasible. To mitigate this issue, we have employed the empirical tight-binding mode space approximation to simulate triple heterojunction TFETs with the body thickness up to 12 nm. The triple heterojunction TFET design is optimized using the model to achieve a sub-60mV/dec transfer characteristic under realistic scattering conditions.  
\end{abstract}

\section{introduction}
The tunneling field-effect transistor (TFET) being able to surpass the Boltzman limit is considered as a promising candidate to replace metal-oxide-semiconductor field-effect transistor (MOSFET) for the future low-power applications \cite{Bohr2011,Memisevic2016,Memisevic2017,Memisevic2018,Mohata2011,
Sant2016,Gonzalez1997,Appenzeller_2004,Appenzeller_2005,Ionescu_2011,Avci2015}. However, the low transmission probability of the band-to-band-tunneling (BTBT) process limits the ON-current. Therefore, the triple heterojunction TFET has been proposed to increase the ON-current by forming two quantum wells in the tunnel junction to decrease tunneling distance and introduce resonant-enhanced tunneling \cite{Long2017,Huang2018_correct}. 

The carrier transport in such devices depends on three factors: 1) the interaction between confined states in quantum wells and propagating states in conduction and valence bands, 2) the BTBT process of confined states in quantum wells, and 3) the scattering effects that thermalize carriers within the quantum well \cite{Long2016_iedm,Kuang_sispad}. Therefore, the accurate atomistic quantum transport simulation, including scattering mechanisms, is the fundamental approach to model such devices \cite{Ameen2017,Huang2017,Junzhe}. The quantum transport simulation is usually conducted by quantum transmitting boundary method (QTBM) \cite{Lent1990,Luisier2009} or non-equilibrium Green's function (NEGF) \cite{datta_2005,datta_2000} using the recursive Green's function (RGF) algorithm  \cite{Lake2004}.

\begin{figure}[!t]
\center
\includegraphics[width=2.7in]{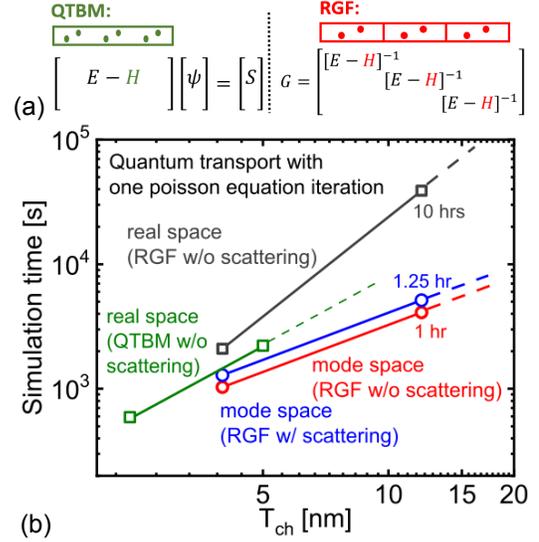}
\caption{(a) The time-consuming part of the two quantum transport algorithms: QTBM and RGF. $H$ and $E$ are the Hamiltonian and energy. $\psi$ and $S$ are the wave function and the strength of the carrier injection from contacts. (b) The quantum transport simulation time for different body thicknesses using the empirical tight-binding basis in the real space and the mode space. The simulations are performed in Nanoelectronics Modeling tool NEMO5 \cite{NEMO5_1,NEMO5_2} by 36 Intel Xeon Gold "Sky Lake" processors.}
\label{simulation_time}
\end{figure}

Both methods are capable of capturing the quantum mechanical effects in nano-devices. However, QTBM cannot capture inelastic scattering. Fig. \ref{simulation_time} (a) shows the schematics of QTBM and RGF. For QTBM, one has to compute the wave functions of an open system. The wave functions are obtained as solutions of the linear system with the size equal to the Hamiltonian matrix dimension. The method is numerically efficient because it computes only a few wave functions per energy, namely, for the states of particles that are injected into the device from the propagating modes in the leads \cite{Lent1990,Luisier2009}. Since the method is based on the wave function formalism, it cannot describe incoherent processes such as inelastic scattering. For RGF, the device is partitioned into thin slabs, and its Green's function is solved recursively. Since the calculation of Green's function requires matrix inversion, RGF is usually slower than QTBM. However, since RGF is capable of including scattering mechanisms, in this work, RGF is chosen to study the triple heterojunction TFET.

The computation cost of RGF is O($N^{3}N_{s}$) where $N_{s}$ is the number of the slab in the device, and $N$ is the matrix size of each slab. $N$ is proportional to the number of atoms per slab times the number of the orbitals per atom \cite{Huang2018_correct}. Since the number of the atoms grows with the device dimension, the computation becomes too expensive for the realistic body thicknesses (i.e. $>$ 8 nm) \cite{Lee_2006,Memisevic2016,Fujimatsu2013,Heterojunction2017,Zhao2017,chinyi2018,
chinyi_2013,Simon2019,xsede_resource}. Fig. \ref{simulation_time} (b) shows how the simulation time of an ultra-thin body (UTB) TFET grows with body thickness. RGF simulation time for 12 nm thick devices using the empirical tight-binding $sp^3d^5s^*$ basis in real space is 10 hours per Poisson equation iteration. 1000 hours are needed to obtain the transfer characteristic of a transistor, considering the number of required iterations and bias points,  which is prohibitively time-consuming for device optimization. 

The mode-space (MS) approximation \cite{Milnikov2012,Huang2018_correct} that compresses the basis to reduce the simulation time becomes
necessary to enable device research for the body thickness exceeding 10 nm. With the MS approximation, the RGF simulation time for a 12 nm thick device is reduced to one hour, which allows studying the device characteristics in detail. 

Previously, the MS approximation has
been used to simulate nanowire MOSFETs and homojunction
UTB TFETs \cite{Huang2018_correct,Jeong2016}. In this work, we expand the method to simulate UTB heterojunction TFETs. To describe the scattering of carriers in quantum wells, an efficient thermalization model, that showed to match experimental data, has been incorporated into the MS approximation \cite{Huang2017,Ameen2017}. The simulation time of a device with 12 nm body thickness increases by 25$\%$ if the scattering is included, which is practically acceptable.

The paper is divided into four sections. In section II, we present the transferable transformation matrix for different transverse wave vectors in UTB applications. In section III, the working principle of the triple heterojunction TFET is discussed. The full empirical tight-binding basis in real space and the mode space are benchmarked for transfer characteristics and local density of states (LDOS). In section IV and V, the performance of the triple heterojunction TFETs with a body thickness of 12 nm is evaluated in the ballistic limit, and the impact of scattering is analyzed.

\begin{figure}[!b]
\center
\includegraphics[width=2.7in]{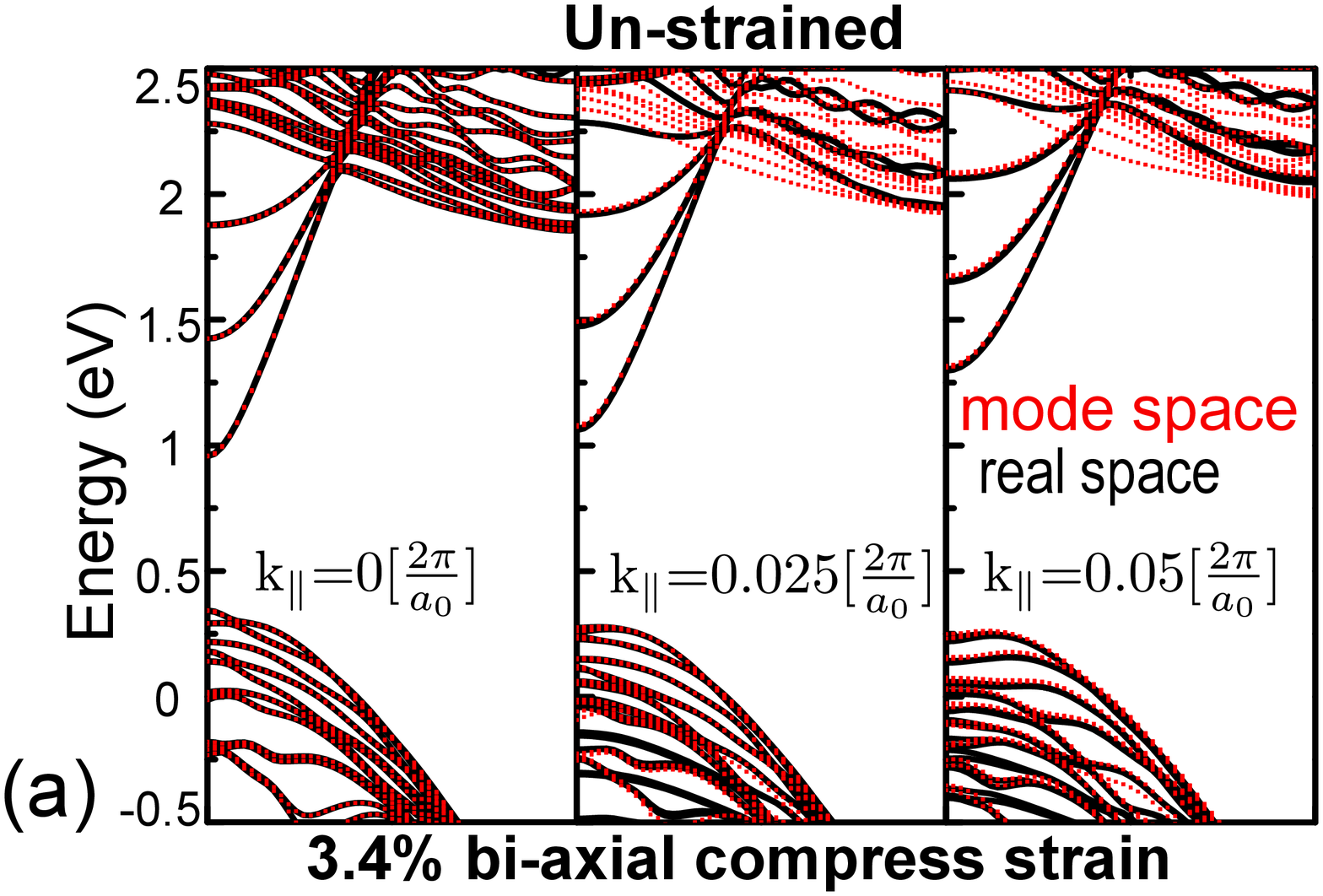}
\includegraphics[width=2.7in]{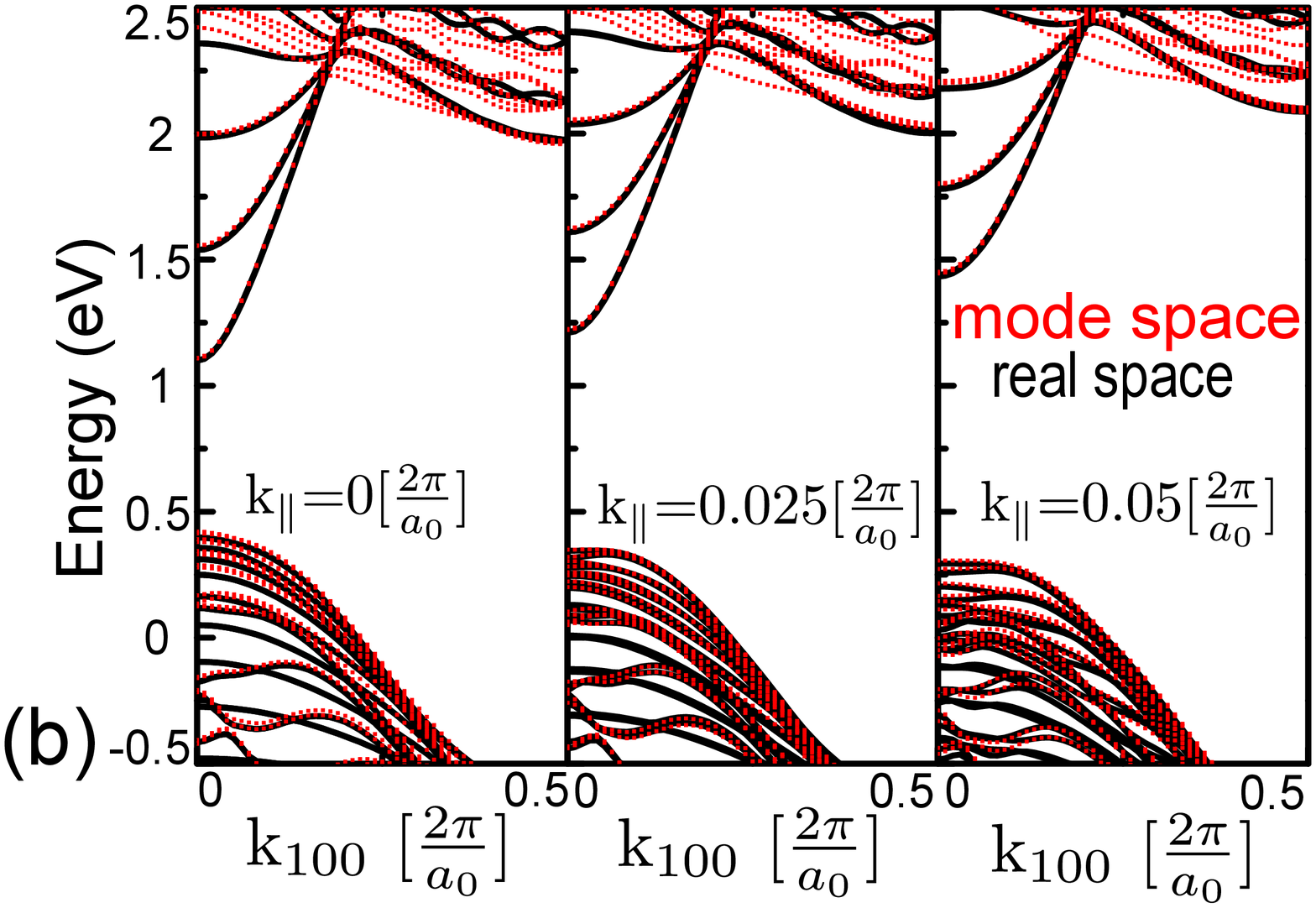}
\caption{Electronic band structure for a 4 nm InAs UTB computed using both empirical tight-binding basis in the real space and the mode space with (a) zero strain and (b) 3.4$\%$ bi-axial compressive strain. The zero strain transformation matrix generated at $k_{\parallel}$= 0.025 $\frac{2\pi}{a_{0}}$ is used to obtain the above mode space basis. The basis reduction ratio ($n/N$) is 178/800.}
\label{fig2}
\end{figure}


\section{Transferable transformation matrix}

For the MS approximation, the Hamiltonian size is reduced by the transformation:  

\begin{align}   
h(k_{\parallel})_{n \times n}=U^{T}_{n \times N}H(k_{\parallel})_{N \times N}U_{N \times n}
\label{eq_reduced_hamiltonian}        
\end{align}


where $H$ is full basis Hamiltonian constructed by the empirical tight-binding (ETB) method and $U$ is the transformation matrix generated by the mode space algorithm by optimizing the modes near the band edges \cite{Milnikov2012,Huang2018_correct}. $h$ is the reduced-size mode space Hamiltonian capturing the modes near the band edges, which contribute significantly to electronic transport. The accuracy of the sub-bands far from the band edges is sacrificed to reduce the representation such that $n$ is significantly smaller than $N$ \cite{Milnikov2012,Huang2018_correct}.

The electronic transport in a UTB system requires sampling multiple transverse wave vectors ($k_{\parallel}$) along the periodic direction. Traditionally, the reduced-size Hamiltonians at each sampled $k_{\parallel}$ are supposed to be generated by different transformation matrices because the modes contributing to electronic transport are different for different $k_{\parallel}$. However, generating the transformation matrix for each sampled $k_{\parallel}$ is a time-consuming process due to the basis optimization \cite{Milnikov2012,Huang2017}. In this work, we find out that generating the transformation matrix for each sampled $k_{\parallel}$ is not necessary since the matrix is transferable within a sizable range of $k_{\parallel}$. 

Fig. \ref{fig2}(a) is the electronic structures of a 4 nm InAs UTB grown along the [100] direction and confined along the [011] direction. The transformation matrix is generated for $k_{\parallel}=$ 0.025 $\frac{2\pi}{a_{0}}$. It can be used to reduce the basis size while capturing the modes nearby the band edges for $k_{\parallel}$ in the range of 0 to 0.05 $\frac{2\pi}{a_{0}}$. However, for $k_{\parallel}$ that is larger than 0.05 $\frac{2\pi}{a_{0}}$, it can not capture the modes nearby the band edges since the modes are too different from those of $k_{\parallel}=$ 0.025 $\frac{2\pi}{a_{0}}$. Considering these facts, one can divide the whole $k_{\parallel}$ space into several segments and use the same transformation matrix to generate the reduced-size Hamiltonian within each segment. This feature is the critical feature for applying MS approximation in a UTB system as it avoids the efforts to generate redundant transformation matrices.


The transformation matrix is not just transferable for different $k_{\parallel}$. It is also transferable for different strain conditions. For heterojunction devices, the strain is a critical and inherent factor that affects material properties like the effective mass and the bandgap. Fig. \ref{fig2}(b) shows the electronic structure of a 4 nm InAs UTB with 3.4$\%$ bi-axial compressive strain, which will be used in the triple heterojunction TFET studied later in this work. In Fig. \ref{fig2}(b), the same transformation matrix has been used as the one in Fig. \ref{fig2}(a). Since the transformation matrix is transferable for different strains, one can easily use MS approximation for devices with different strain conditions.



%


\section{Method validation}
\begin{figure}[!t]
\includegraphics[width=1.6in]{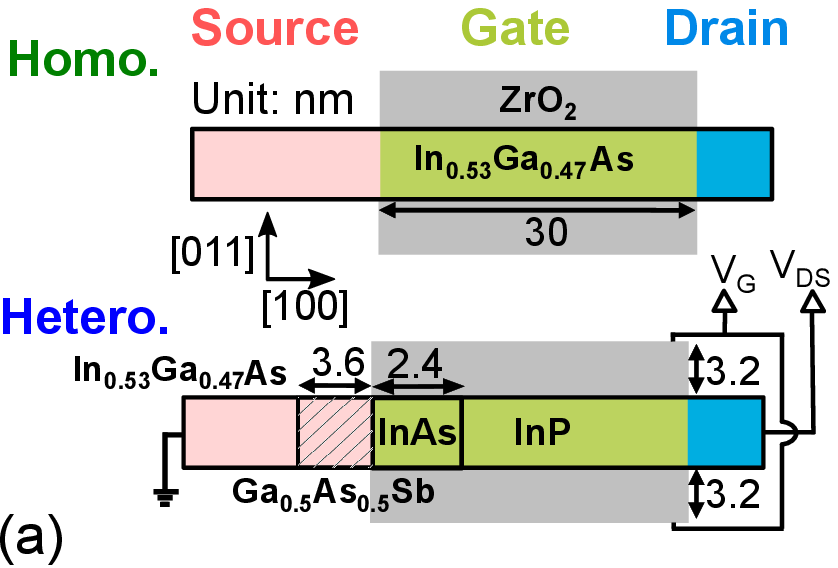}
\includegraphics[width=1.6in]{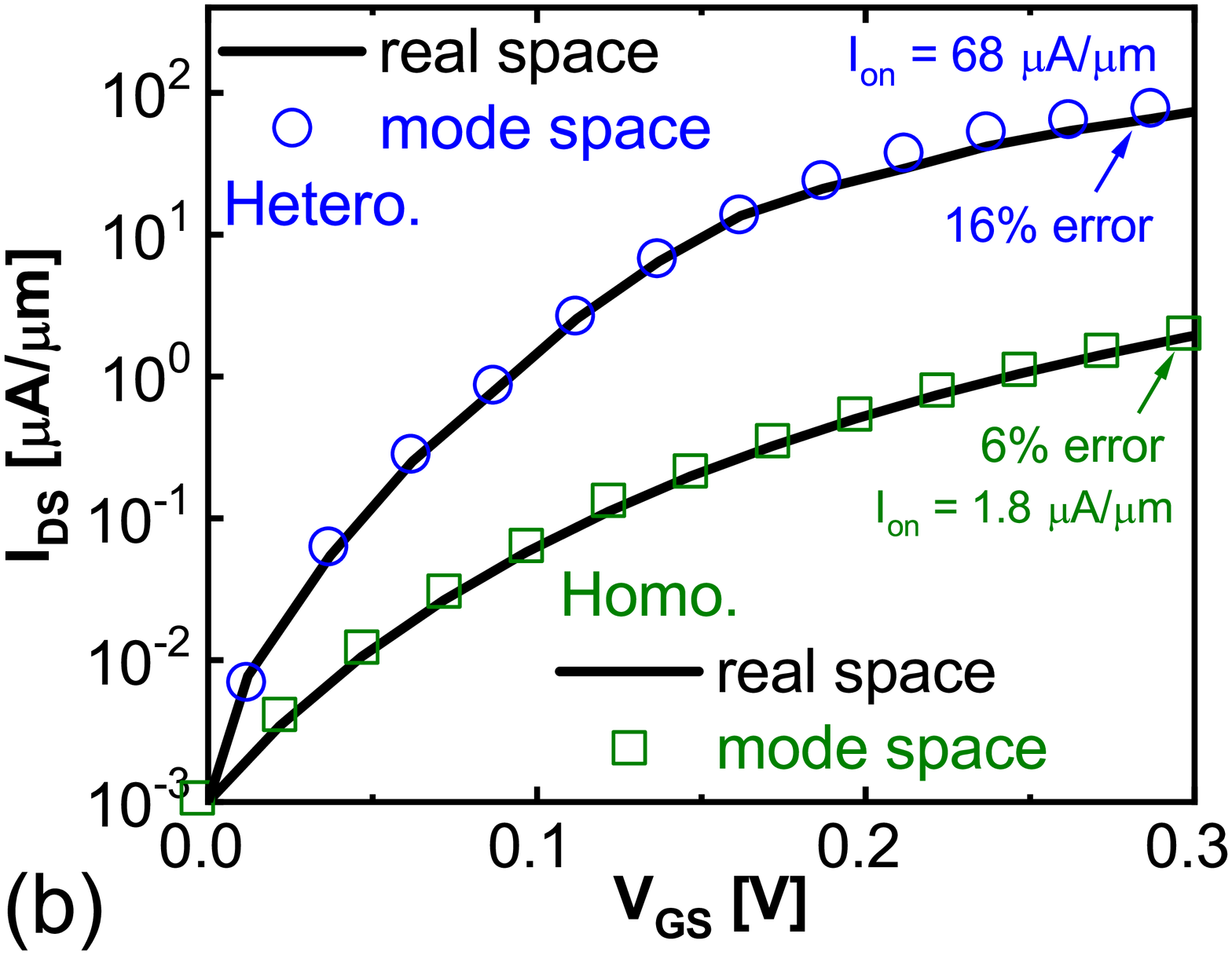}
\includegraphics[width=1.6in]{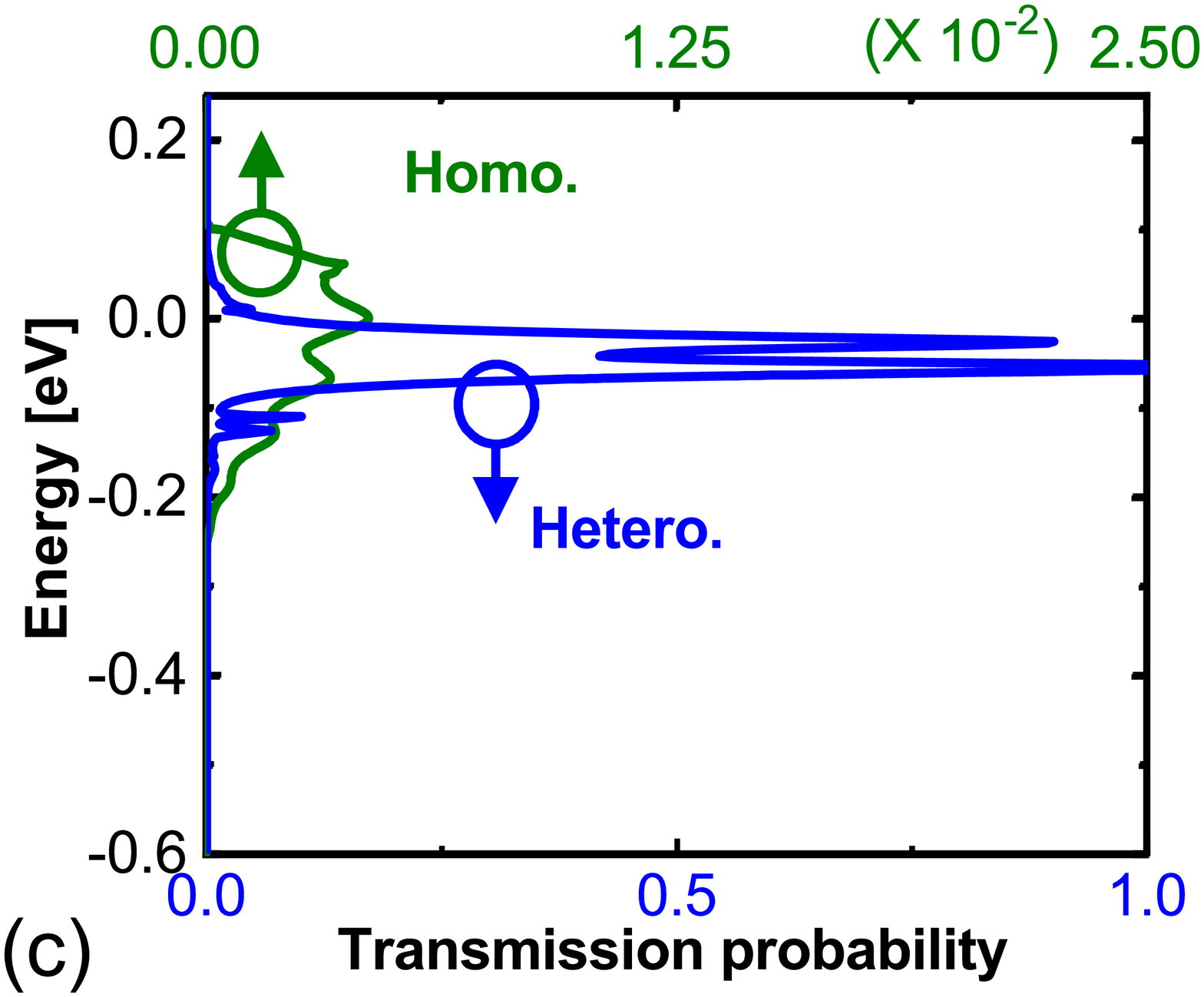}
\includegraphics[width=1.6in]{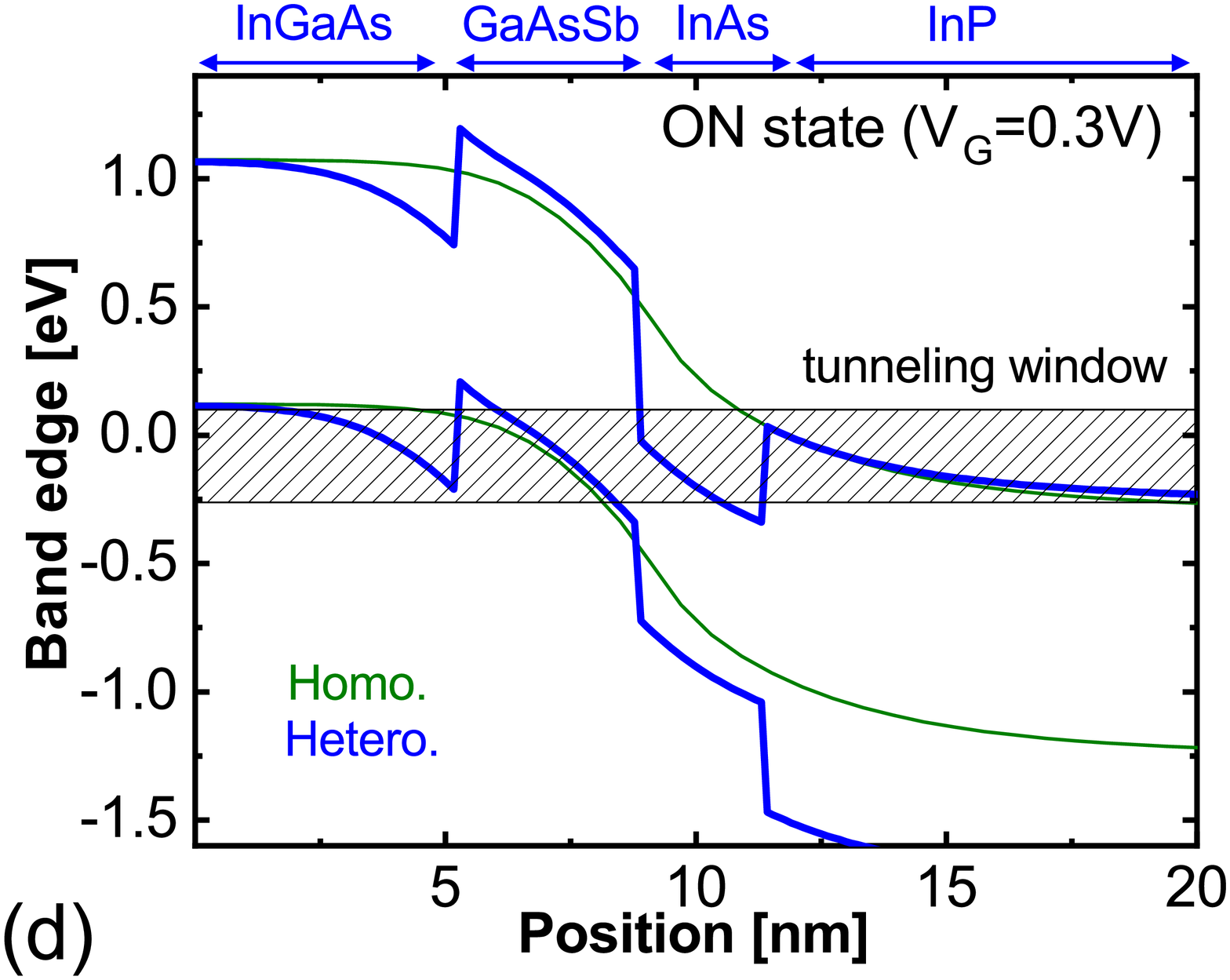}
\caption{ (a) Device schematics, (b) transfer IV characteristic, (c) transmission probability at $V_{GS}$ = 0.3 V, and (d) band diagram at $V_{GS}$ = 0.3 V of double-gate InGaAs homojunction and triple heterojunction TFET with 4 nm body thickness. The transfer IV characteristic is computed using the empirical tight-binding basis in both the real space and the mode space.  }
\label{schematic}
\end{figure}


In this section, the MS approximation is validated for the In$_{0.53}$Ga$_{0.47}$As homojunction (homo-) and the triple heterojunction (hetero-) TFETs. The double-gate UTB schematics are shown in Fig. \ref{schematic} (a). The body thickness of 4 nm is chosen since using the full basis to simulate a complete I-V curve is computationally expensive as the body thickness increases further. The triple heterojunction TFET design consists of a P-doped In$_{0.53}$Ga$_{0.47}$As and GaAs$_{0.5}$Sb$_{0.5}$ source with $N_{a}=5 \times 10^{19}$ $cm^{-3}$, an intrinsic InAs and InP channel, and an N-doped InP drain with $N_{d}=2 \times 10^{19}$ $cm^{-3}$. The confinement direction is along the [011] direction, and the transport direction is along the [100] direction. The crystal growth direction is along the transport direction to simulate the vertical Fin-TFET structure \cite{Fujimatsu2013}. The substrate is assumed to be InP such that InAs is under 3.41$\%$ bi-axial compressive strain while the rest of the materials are not under strain. The gate dielectric is a 3.2 nm thick ZrO$_{2}$ with the relative dielectric constant of 15. The source to drain bias ($V_{DS}$) is 0.3 V.


The InGaAs homojunction and triple heterojunction TFET's transfer IV characteristic ($I_{DS}$-V$_{GS}$) are shown in Fig. \ref{schematic} (b). The IV curves are shifted to have a fixed OFF-current value of $10^{-3}$ $\mu A/\mu m$ at $V_{GS}$ = 0 V. The current obtained from the MS approximation agrees with the current calculated by the full ETB basis. The error introduced by the MS approximation is quantified through the expression $\Delta I_{err}$=$| I_{full}-I_{MS}|/I_{full}$. At the ON-state, where $V_{G}$ = 0.3 V, the triple heterojunction TFET's $\Delta I_{err}$ is 16$\%$, which is slightly higher than the InGaAs homojunciton TFET's $\Delta I_{err}$ which is 6$\%$. The higher error in the heterojunction case is expected due to the presence of junction interfaces, and more transformation matrices are used (one for each material). 


The ON-current ($I_{ON}$) of the triple heterojunction TFET is 68 $ \mu A/\mu m$, which is much higher than the InGaAs homojunction TFET's $I_{ON}$ 1.8 $\mu A/\mu m$. The reason is that the triple heterojunction TFET's tunneling distance is smaller than the InGaAs homojunction TFET's tunneling distance due to the staggered-heterojunction (GaAsSb/InAs) used in BTBT tunneling region as shown in Fig. \ref{schematic} (d). Since InGaAs and InP are present in the source and the channel, two quantum wells are formed in the tunneling junction. The quantum well states introduce resonant-enhanced tunneling that boosts the transmission probability close to 1, as shown in Fig. \ref{schematic} (c).

\begin{figure}[!t]
\center
\includegraphics[width=3.4in]{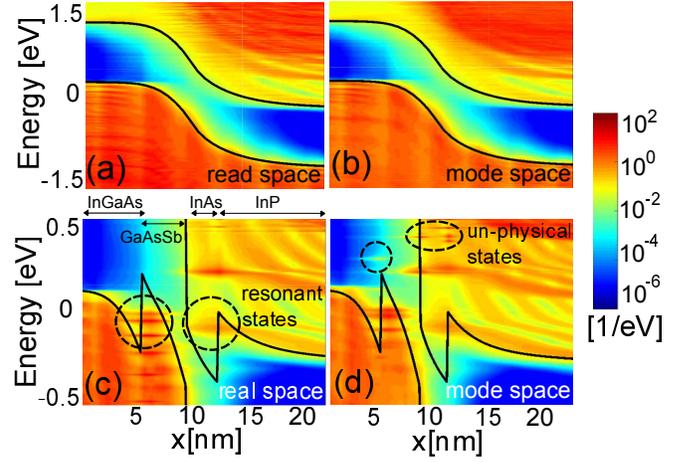}
\caption{LDOS of (a),(b) the InGaAs homojunction TFET and (c),(d) the triple heterojunction TFET at the ON-state (V$_{GS}$ = 0.3 V) along the channel computed using the empirical tight-binding basis in both the real space and the mode space.}
\label{4nm_LDOS_compare}
\end{figure}


To gain a better understanding of the MS approximation, the local density of states (LDOS) of the InGaAs homojunction and triple heterojunction TFET at the ON-state ($V_{GS}$= 0.3 V) are shown in Fig. \ref{4nm_LDOS_compare}. In (a) and (b), the InGaAs homojunction TFET's LDOS computed from the real space and the mode space are almost the same. While in (c) and (d), the triple heterojunction TFET's LDOS computed from the real space and mode space has some differences. Aside from the aligned resonant states that introduce the resonant-enhanced tunneling, some extra un-physical states at the heterojunction interfaces are visible in the LDOS computed by the mode space. The resonant states and the un-physical states are marked with black dashed circles. 


The material interface needs extra attention in the MS approximation since the coupling Hamiltonian at the interface is multiplied with the transformation matrices of both materials. That's why we see the un-physical localized states near the interfaces. These un-physical localized states are part of the reason why the triple heterojunction TFET's $\Delta I_{err}$ is higher than the InGaAs homojunction TFET's $\Delta I_{err}$.



\begin{figure}[!t]
\center
\includegraphics[width=2.5in]{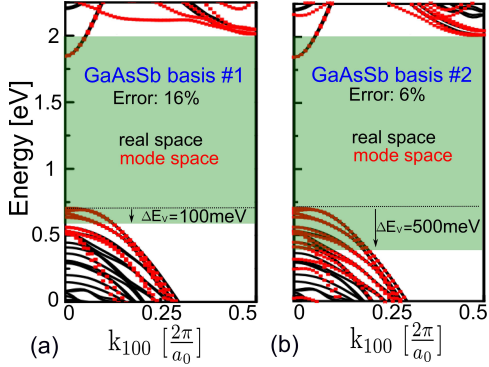}
\caption{Electronic structure of a 4 nm GaAsSb UTB with the energy window that covers the valence band ($\Delta$E$_{V}$) with (a) 100 meV and (b) 500 meV. The shading shows the energy window used for the bases optimization. The basis reduction ratios (n/N) are (a) 125/800 and (b) 200/800.    }
\label{GaAsSb_MS_basis}
\end{figure}

The energy window of the MS basis for materials used in the tunneling junction (GaAsSb and InAs) should be increased to reduce triple heterojunction TFET's $\Delta I_{err}$. Fig. \ref{GaAsSb_MS_basis} shows the basis optimization with different energy windows for the 4 nm GaAsSb UTB used in the triple heterojunction TFET. GaAsSb basis $\#$1 has the energy window that covers the valence band ($\Delta$E$_{V}$) for 100 meV, which is used to obtain the results shown previously in Fig. \ref{schematic}. GaAsSb basis $\#$2's $\Delta$E$_{V}$ is 500 meV, which is larger than the quantum well depth of 420 meV in the triple heterojunction TFET. When GaAsSb basis $\#$1 is replaced with GaAsSb basis $\#$2, the error is reduced from 16 $\%$ to 6$\%$. Since the accuracy of the confined states is critical to electronic transport in the triple heterojunction TFET, the MS approximation for such applications requires a large enough energy window, which covers the depth of the quantum well. 




\section{12 nm body thickness triple heterojunction TFETs}

\begin{figure}[!b]
\center
\includegraphics[width=1.6in]{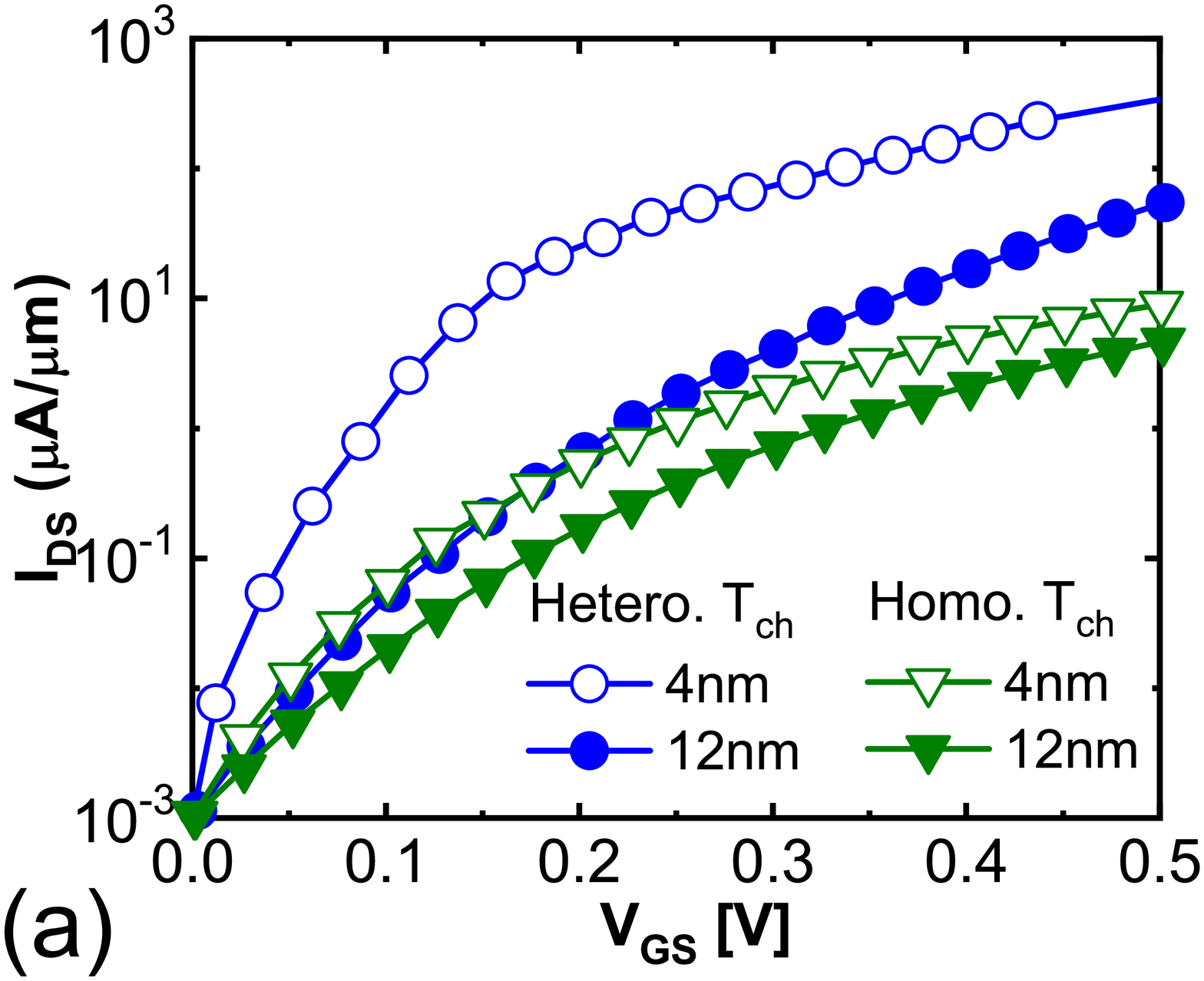}
\includegraphics[width=1.6in]{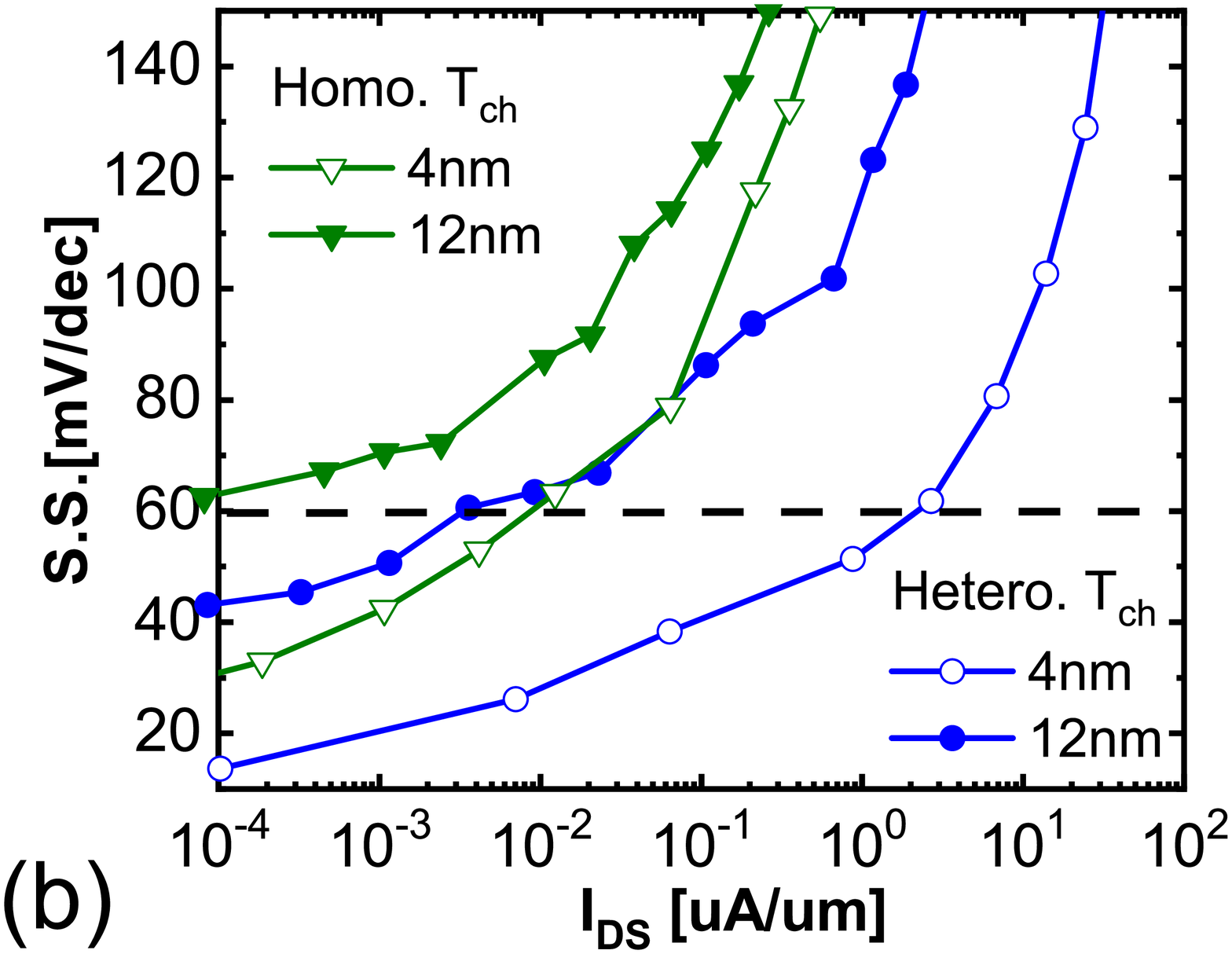}
\caption{(a) Transfer IV characteristic and (b) sub-threshold swing (S.S.) for InGaAs homojunction and triple heterojunction TFET with the body thickness ($T_{ch}$) of 4 nm and 12 nm.}
\label{12nm_tch_IV}
\end{figure}
\begin{figure}[!b]
\center



\includegraphics[width=3.5in]{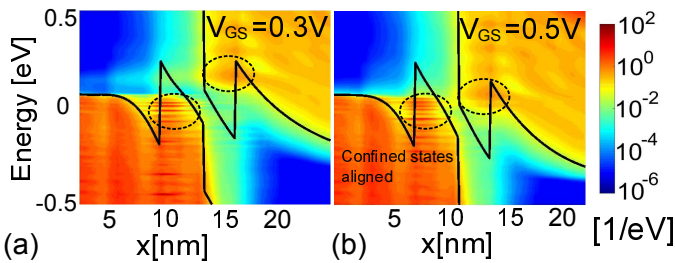}
\caption{LDOS along the channel of triple heterojunction TFET with a body thickness of 12 nm at (a) V$_{GS}$ = 0.3 V and (b) V$_{GS}$ = 0.5 V. The quasi-bound states of the quantum states are indicated by the dashed lines.}
\label{12nm_tch_LDOS}
\end{figure}




The reported triple heterojunction TFET designs in the references \cite{Long2017,Huang2018_correct} were optimized for the body thickness of 4 nm due to the computational limits. However, devices with such a thin body are difficult to be realized in experiments. With the MS approximation, we can increase the simulated body thickness to 12 nm, which is the thinnest possible body thickness for III-V materials Fin-TFET structure \cite{Simon2019}. 

The transfer characteristics and the extracted sub-threshold swing (S.S.) of the triple heterojunction TFET with the body thickness of 12 nm are shown in Fig. \ref{12nm_tch_IV} (a) and (b). The InGaAs homojunction TFET's simulation results are also displayed as a reference. When the body thickness increases, the triple heterojunction TFET's I$_{ON}$ decreases significantly while the InGaAs homojunction TFET's $I_{ON}$ does not. Nevertheless, the triple heterojunction TFET's sub-threshold swing retains sub-60mV/dec which is still better than the InGaAs homojunction TFET.  

The triple heterojunction TFET's I$_{ON}$ depends on the alignment of the resonant states in the quantum wells. Table. ~\ref{table:12nm_Ion} summarizes $I_{ON}$ of the InGaAs homojunction and triple heterojunction TFET with 12 nm body thickness at $V_{GS}$ = 0.3 V and 0.5 V. For the InGaAs homojunction TFET, $I_{ON}$ at $V_{GS}$ = 0.3 V or 0.5 V are of the same order. The impact of the gate bias is limited due to the weak gate control and, consequently, a large scaling length \cite{chinyi2018,Ilatikhameneh2017_sensitive}. In contract with the InGaAs homojunction TFET, the triple heterojunction TFET has the resonant enhanced tunneling that compensates the enlarged tunneling distance due to the loose gate control. Once $V_{GS}$ is large enough to achieve the resonant enhanced tunneling condition, $I_{ON}$ increases from 3.9 $\mu$A/$\mu$m to 50 $\mu$A/$\mu$m. Fig. \ref{12nm_tch_LDOS} (a) and (b) show the LDOS of the triple heterojunction TFET with 12 nm body thickness computed at $V_{GS}$ = 0.3 and 0.5 V. For $V_{GS}$ = 0.5 V, the two resonant states are aligned to achieve the resonant enhanced tunneling condition such that the $I_{ON}$ increases up to 50 $\mu$A/$\mu$m.  

\begin{table} [!b]
\begin{center}
\rule{0pt}{8pt}
\begin{tabular}{ c c c  } 
 \hline
 \hline
 & &    \\ [-0.5em]
$I_{ON}$[$\mu A/\mu m$] & Homo.     & Hetero. \\ 
\hline
 & & \\ [-1em]
 $V_{GS}$=0.3V & 0.7  & 3.9                \\

  & &  \\ [-1em]
 $V_{GS}$=0.5V  & 4  & 50                \\
 \hline
 \hline
\end{tabular}
\end{center}
\caption{The I$_{ON}$ for InGaAs homojunction (homo.) and triple heterojunction (hetero.) TFETs with $T_{ch}$=12 nm.}
\label{table:12nm_Ion}
\end{table}

\section{Efficient scattering model}
The strong scattering and thermalization in the highly doped source and drain regions have a significant impact on the transport properties of tunneling devices \cite{klimeck1995,Klimeck1994,Lake1997,Long2016_drc,Long2016_iciprm,Huang2016_3HJ_TFET,Long2016_iedm}. Different mechanisms such as electron-electron scattering \cite{Archana_2018,mazzola2019subband,tankasala2015atomistic,tankasala2017engineering
,hsueh2016phonon,salfi2018valley,wang2016highly,Wang2016_quantom_dot}, electron-phonon scattering, electron-ion scattering, plasmon scattering, etc. contribute to the strong scattering and thermal ionization. Including all of these scattering mechanism, especially electron-electron scattering, into RGF is computationally unfeasible for realistic devices. Hence, an effective carrier thermalization method is needed. An effective thermalization approach for tunneling devices has been developed for resonant-enhanced tunneling diodes \cite{klimeck1995} and has been shown to match experimental data on Nitride tunneling devices for a wide range of bias conditions \cite{Ameen2017}. In this work, a combination of the mode-space approximation and the thermalization approach is used to include thermalization into atomistic simulation of devices with large and realistic dimensions. 

Thermalization of carriers in source and drain contacts, has two main impacts which significantly distort the transport properties: 1) filling quantum well states near the source region, 2) widening the resonance energies inside the well. Including scattering accurately is crucial, since the states in the quantum well contribute to a significant part of the channel leakage.

In this work, the effective scattering rate in the source and the drain is estimated from the mobility, that empirically represents the strength of the scattering. Since the mobility ($\mu$) of the highly doped III-V materials ranges from $10^{2}$ to $10^{3}$ $cm^{2}/(V.s)$ \cite{Mart_2004,Karishy2015,Krivec2016} with the effective mass ($m^{*}$) of 0.001 $\sim$ 0.01, the reasonable energy broadening ($\eta \sim \frac{q \hbar}{2 m^{*} \mu}$) is about 1 $\sim$ 10 meV. In this work, the broadening factor of 10 meV is used to explore the worst-case scenario \cite{Ameen2017}. The momentum relaxation time is 32 fs ($\tau=\frac{\hbar}{2\eta}$).



\begin{figure}[!t]
\center
\includegraphics[width=3.2in]{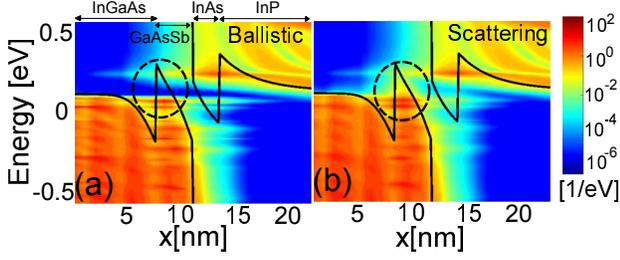}
\caption{LDOS for the triple heterojunction TFET with 4 nm body thickness at the OFF-state (a) with and (b) without scattering effects. The dashed line marks a quantum well region filled by electrons due to the scattering.}
\label{eqneq_LDOS_full}
\end{figure}

For triple heterojunction TFET, the quantum well states in the tunneling junction contribute to the OFF-state leakage current. The OFF-state LDOS of the triple heterojunction TFET with 4 nm body thickness, computed by the ballistic and the scattering model, is shown in Fig. \ref{eqneq_LDOS_full} (a) and (b). The deep quantum well states are marked with the black circle. In the ballistic simulation, the deep quantum well states are slightly occupied by the carriers injected from the contacts. While for the scattering model, the quantum well states are populated by the scattering thermalized carriers, which is close to the real situation when devices operate at room temperature. 

\begin{figure}[!t]
\center
\includegraphics[width=2.0in]{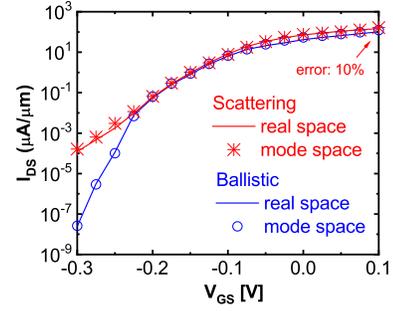}
\caption{Transfer IV characteristics of a triple heterojunction TFET with 4 nm body thickness computed with and without scattering effects considered. }
\label{eqneq_benchmark}
\end{figure}

The thermalized quantum well states introduce the leakage path and increase the sub-threshold leakage. Fig. \ref{eqneq_benchmark} compares the 4 nm thick triple heterojunction TFET's transfer IV characteristics with and without scattering. With scattering, the sub-threshold current at V$_{GS}$ = -0.3 V is $10^{3}$ times higher than the results of the ballistic simulation. The MS approximation is also applied to the scattering simulation, and it introduces a small $\Delta I_{err}$ of 10$\%$.

\begin{figure}[!h]
\center
\includegraphics[width=1.6in]{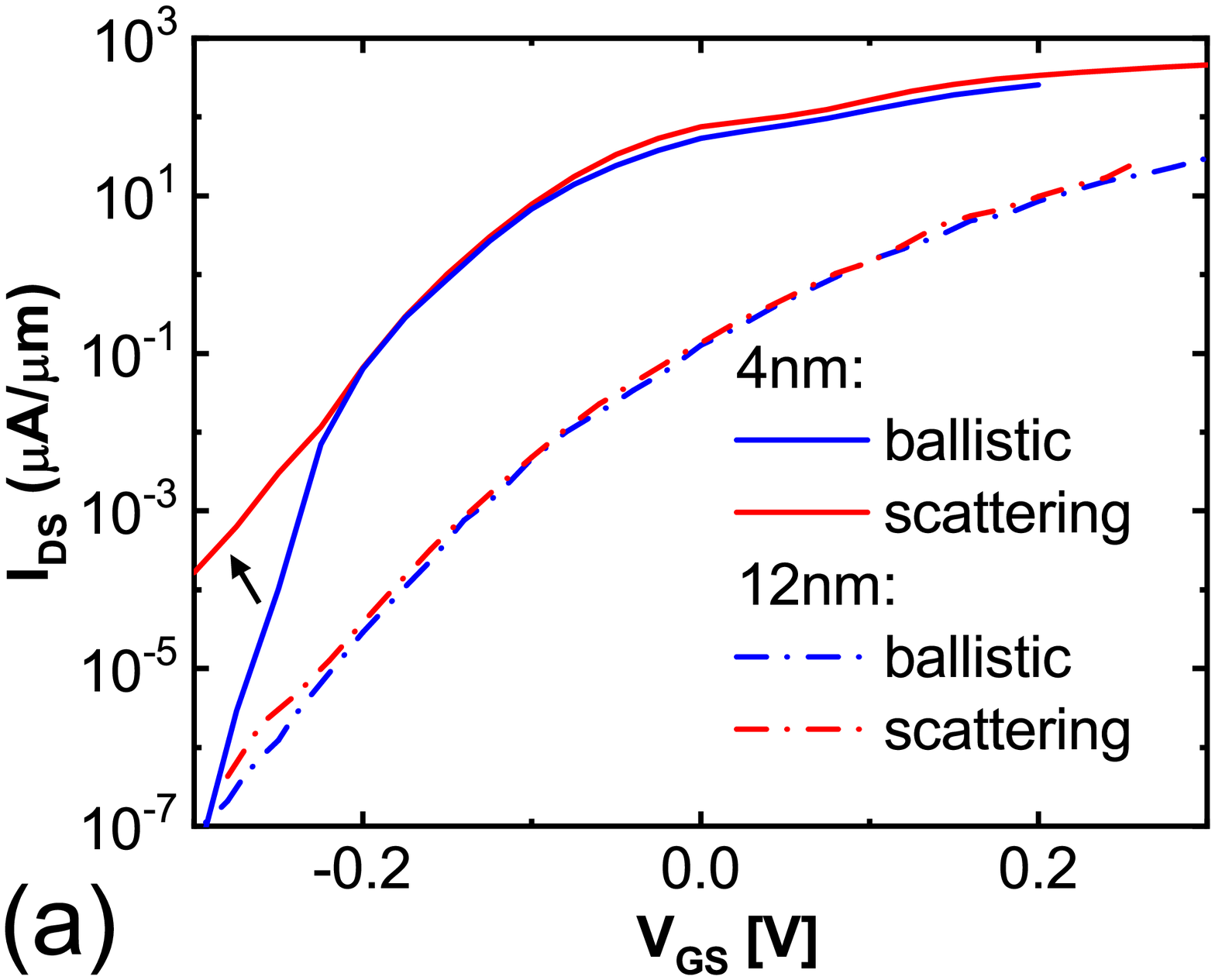}
\includegraphics[width=1.6	in]{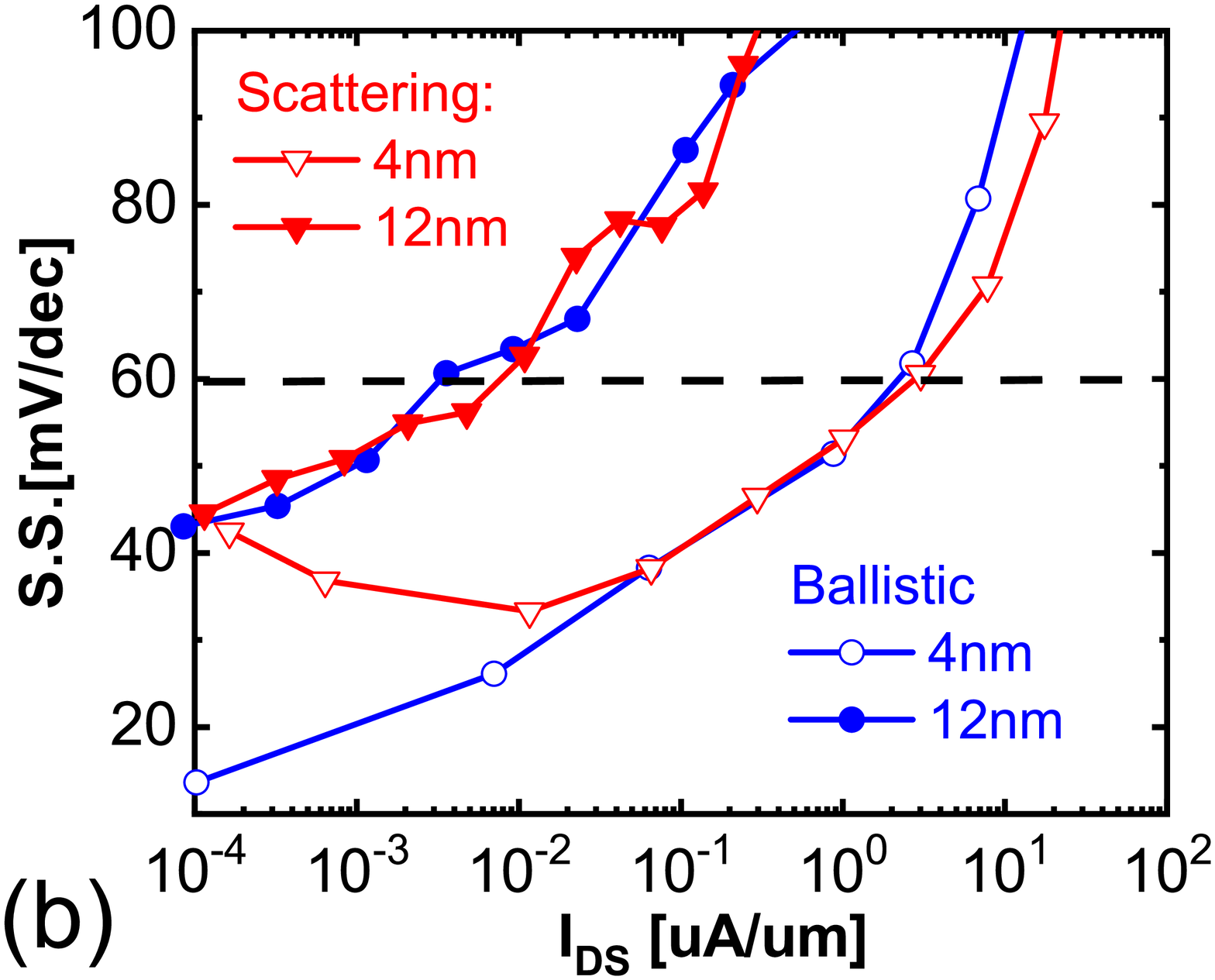}
\caption{ (a) Transfer IV characteristics and (b) sub-threshold swing computed with and without scattering effects for triple heterojunction TFET with the body thickness of 4 nm and 12 nm. }
\label{eqneq_12nm}
\end{figure}

\begin{figure}[!h]
\center
\includegraphics[width=3.5in]{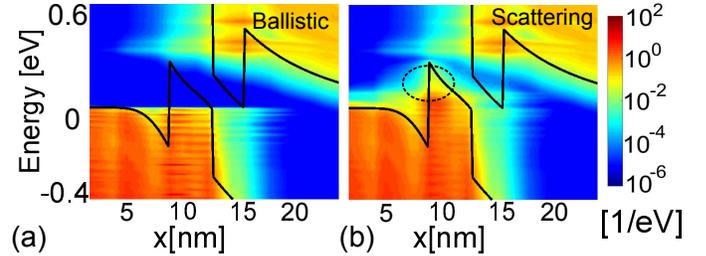}
\caption{LDOS for a triple heterojunction TFET with the body thickness of 12 nm at OFF-state (V$_{GS}$ = -0.3 V) computed from (a) ballistic simulation and (b) simulation with scattering effects. The LDOS contributions due to the scattering effects are indicated with the dashed lines.}
\label{eqneq_12nm_LDOS}
\end{figure}

The results of the scattering model for the 12 nm thick triple heterojunction TFET are shown in Fig. \ref{eqneq_12nm}, where (a) are the transfer IV characteristics, and (b) are the extracted sub-threshold swings. Both the ballistic and the
scattering results are plotted to show the introduced degradation from scattering. Contrary to the 4 nm thickness case, for the 12 nm thickness case, the scattering and the ballistic simulations give very similar results. The reason is that the thermalized quantum well states in the 12 nm thick device have a long tunneling distance in the channel to tunnel through, which leads to a low transmission probability. Fig. \ref{eqneq_12nm_LDOS} (a) and (b) are the 12 nm thick triple heterojunction TFET’s OFF-state LDOS computed by the ballistic model and the scattering model. The marked thermalized quantum well states have a negligible contribution to the OFF leakage due to the long tunneling distance toward the drain.




\section{Summary}
In this work, the triple heterojunction tunneling FinFET with the body thickness of 12 nm is studied using the mode 
space approximation. The sub-threshold swing retains sub-60 mV/dec value, and the degradation due to the scattering is negligible. The transformation matrix that generates the mode space basis is found transferable for different strains and transverse wave vectors, which is a convenient feature for the mode space UTB simulations. When the mode space approximation is applied in the heterojunction devices, un-physical states at the heterojunction interface can be introduced if the mode space energy window is chosen too small. However, the error can be reduced by increasing the mode space basis's energy window to cover the depth of the quantum well. Overall, the combination of the mode space approximation and the empirical scatting model made the analysis of TFET with a realistic dimension possible.

\bibliographystyle{ieeetr}
\bibliography{citation_file}

\end{document}